\documentclass[a4paper,twocolumn,10pt,accepted=2024-05-12]{quantumarticle}
\pdfoutput=1
\RequirePackage{silence}
\RequirePackage{float}
\usepackage[utf8]{inputenc}
\usepackage[USenglish]{babel}
\usepackage[T1]{fontenc}
\usepackage{graphicx,epsf}
\usepackage[toc,page,header]{appendix}
\usepackage{minitoc}
\usepackage{subcaption}
\usepackage{color}
\usepackage{breakurl}
\usepackage[numbers,sort&compress]{natbib}
\usepackage[bookmarksnumbered,hypertexnames=false,colorlinks,colorlinks=true,linkcolor=blue,urlcolor=teal,citecolor=blue,anchorcolor=green]{hyperref}
\usepackage{bbm,braket,microtype,mathrsfs,amsmath,amssymb,color,amsthm,mathtools,fullpage,enumitem,bm,thm-restate}
\usepackage{tikz}
\usepackage{dsfont}
\usepackage{nameref}
\usepackage{caption}
\usepackage{color}
\usepackage{comment}
\usepackage[capitalize]{cleveref}
\usepackage{silence}
\allowdisplaybreaks
\newcommand{\be}{\begin{equation}}
\newcommand{\ee}{\end{equation}}
\newcommand{\ba}{\begin{eqnarray}}

\newcommand{\ea}{\end{eqnarray}}
\DeclareMathOperator{\tr}{tr}

\newcommand{\ignore}[1]{}
\newcommand{\st}[1]{\ket{#1}\!\!\bra{#1}}

\newcommand{\haar}[0]{\operatorname{Haar}}

\newcommand{\poly}{\operatorname{poly}}

\newcommand{\id}{\mathds{1}}

\newtheorem{algorithm}{Algorithm}
\def\CC{{\rm\kern.24em \vrule width.04em height1.46ex depth-.07ex
   \kern-.29em C}}
\def\P{{\rm I\kern-.25em P}}
\def\RR{{\rm
        \vrule width.04em height1.58ex depth-.0ex
        \kern-.04em R}}
\def\id{{\mathchoice {\rm 1\mskip-4mu l} {\rm 1\mskip-4mu l}
{\rm 1\mskip-4.5mu l} {\rm 1\mskip-5mu l}}}
\def\bbbc{{\mathchoice {\setbox0=\hbox{$\displaystyle\rm C$}\hbox{\hbox
to0pt{\kern0.4\wd0\vrule height0.9\ht0\hss}\box0}}
{\setbox0=\hbox{$\textstyle\rm C$}\hbox{\hbox
to0pt{\kern0.4\wd0\vrule height0.9\ht0\hss}\box0}}
{\setbox0=\hbox{$\scriptstyle\rm C$}\hbox{\hbox
to0pt{\kern0.4\wd0\vrule height0.9\ht0\hss}\box0}}
{\setbox0=\hbox{$\scriptscriptstyle\rm C$}\hbox{\hbox
to0pt{\kern0.4\wd0\vrule height0.9\ht0\hss}\box0}}}}
\def\bbbz{{\mathchoice {\hbox{$\sf\textstyle Z\kern-0.4em Z$}}
{\hbox{$\sf\textstyle Z\kern-0.4em Z$}}
{\hbox{$\sf\scriptstyle Z\kern-0.3em Z$}}
{\hbox{$\sf\scriptscriptstyle Z\kern-0.2em Z$}}}}

\newlength{\fighskip} \fighskip=2pt
\newlength{\figvskip} \figvskip=1pt

\makeatletter
\def\namedlabel#1#2{\begingroup
   \def\@currentlabel{#2}%
   \label{#1}\endgroup
}
\makeatother

\usepackage{hyperref}

\newtheorem{lemma}{Lemma}

\definecolor{KB}{rgb}{0.4,0.3,0.9}
\begin{document}
\setcounter{secnumdepth}{3}
%\onecolumngrid
\title{Learning $t$-doped stabilizer states}

\author{Lorenzo Leone}\email{lorenzo.leone@fu-berlin.de}
\affiliation{Physics Department,  University of Massachusetts Boston,  02125, USA}
\affiliation{Dahlem Center for Complex Quantum Systems, Freie Universit\"at Berlin, 14195 Berlin, Germany}

\author{Salvatore F.E. Oliviero}\email{salvatore.oliviero@sns.it}
\affiliation{Physics Department,  University of Massachusetts Boston,  02125, USA}

\affiliation{NEST, Scuola Normale Superiore and Istituto Nanoscienze, Consiglio Nazionale delle Ricerche, Piazza dei Cavalieri 7, IT-56126 Pisa, Italy}

\author{Alioscia Hamma}\email{alioscia.hamma@unina.it}

\affiliation{Dipartimento di Fisica `Ettore Pancini', Universit\`a degli Studi di Napoli Federico II,
Via Cintia 80126,  Napoli, Italy}
\affiliation{INFN, Sezione di Napoli, Italy}
%\affiliation{Physics Department,  University of Massachusetts Boston,  02125, USA}

\begin{abstract}
In this paper, we present a learning algorithm aimed at learning states obtained from computational basis states by Clifford circuits doped with a finite number $t$ of $T$-gates. The algorithm learns an exact tomographic description of $t$-doped stabilizer states in terms of Pauli observables. This is possible because such states are countable and form a discrete set. To tackle the problem, we introduce a novel algebraic framework for $t$-doped stabilizer states, which extends beyond $T$-gates and includes doping with any kind of local non-Clifford gate. The algorithm requires resources of complexity $\operatorname{poly}(n,2^t)$ and exhibits an exponentially small probability of failure.
\end{abstract}

\maketitle

\section{Introduction} The task of learning an unknown quantum state usually refers to the ability to build a classical representation of the state that is ``close'' enough, based on a specified measure of distance, to the unknown state. However, it is well known that learning an arbitrary many-body quantum state requires an exponential number of experiments due to the exponential growth of the Hilbert space with the number of particles.  

Let us be more precise, let us consider the Hilbert space of $n$ qubits and denote $d\equiv 2^n$ its dimension. We can define the group of Pauli operators as the collection of all possible products of Pauli matrices applied to each qubit. Pauli tomography~\cite{paris_quantum_2004,cramer_efficient_2010,gross_quantum_2010,aaronson_shadow_2018,huang_classical_2020,huang_predicting_2020} serves as a prevalent and widely employed technique for learning a given state $\psi$. Specifically, Pauli tomography entails acquiring knowledge about the expectation values $\tr(P\psi)$ for every Pauli string $P$. To achieve a classical reconstruction of the state vector $\psi$, a total of $d^2$ Pauli operators must be measured and it becomes evident that tomography is generally a resource-intensive task. The primary challenge in achieving efficiency in quantum state tomography arises from our limited knowledge of the state of interest. 
Nevertheless, there exist different classes of states, for which tomography can be performed with remarkable efficiency, e.g. Matrix Product States~\cite{cramer_efficient_2010}, quantum phase states~\cite{arunachalam_optimal_2023}, non-interacting fermionic states~\cite{aaronson_efficient_2023} and stabilizer states~\cite{gottesman_heisenberg_1998}. The latter is obtained by the computational basis state $\ket{0}^{\otimes n}$ by the action of the Clifford group -- the subgroup of the unitary group that sends Pauli operators in Pauli operators~\cite{gottesman_heisenberg_1998}. Each stabilizer state $\ket{\sigma}$ corresponds to a stabilizer group $G_{\sigma}$, a subgroup of the Pauli group, comprising $d$ mutually commuting Pauli operators $P$ satisfying the condition $P\ket{\sigma}=\pm \ket{\sigma}$, i.e. elements of the stabilizer group $G_{\sigma}$ stabilize the stabilizer state $\ket{\sigma}$. Thus, to find the tomographic description of $\sigma$, one needs to learn the stabilizer group $G_{\sigma}$ and the relative phases. Every stabilizer group $G_{\sigma}$ admits (more than) a set of generators $g_{\sigma}$, i.e. $G_{\sigma}=\langle g_{\sigma}\rangle$ with cardinality $|g_{\sigma}|=n$. Thus, it is sufficient to find a set of $n$ generators and phases to completely characterize $\sigma$. Montanaro~\cite{montanaro_learning_2017} showed an algorithm that learns the tomographic decomposition of any stabilizer state with $O(n)$ queries to $\ket{\sigma}$, thus gaining an exponential speed-up for stabilizer states overall tomographic methods~\cite{bruss_optimal_1999}. 

However, stabilizer states are not universal for quantum computation and cannot provide quantum speed-up of any kind~\cite{gottesman_theory_1998}. To make the set of stabilizer states universal for quantum computation, one must use unitaries lying outside the Clifford group~\cite{gottesman_theory_1998}. One common choice is to use the single qubit $T$-gate defined as $T=\operatorname{diag}(1,e^{i\pi/4})$. States obtained from $\ket{0}^{\otimes n}$ by the action of Clifford unitaries plus $t$ $T$-gates are usually referred to as $t$-doped stabilizer states~\cite{leone_quantum_2021,oliviero_transitions_2021}.

Whether Montanaro's procedure could be extended to states beyond the stabilizer formalism, in particular to $t$-doped stabilizer states, has been a research question since then. There has been a noteworthy attempt in~\cite{lai_learning_2022}, in which the authors cleverly discovered an extension of the stabilizer formalism for those states obtained by a stabilizer state acting with a row of $t$ $T$-gates followed by a Clifford circuit: \textit{T-depth $1$} circuits. Such states can be learned by $O(3^tn)$ query accesses and time $O(n^3+3^tn)$. However, their protocol is limited only to those states achievable through a specific architecture of Clifford$+T$ circuits.

In this paper, we finally show that it is possible to extend learning algorithms to general states obtained by Clifford$+T$ unitaries. We propose an innovative algebraic framework for $t$-doped stabilizer states by employing concepts from stabilizer entropy~\cite{leone_stabilizer_2022}. Building upon this newly established structure, we devise two algorithms that aim at learning an unknown $t$-doped stabilizer state exactly. The first one involves sampling from a distribution $\Xi$ obtained by squaring the expectation values of Pauli operators, also known as \textit{characteristic distribution}, that can be achieved by a Bell sampling on both the state and its conjugate in the computational basis~\cite{montanaro_learning_2017}. The second approach relaxes the requirement of accessing samples from the characteristic distribution. Instead, it relies on the sampling from a probability distribution $\tilde{\Xi}$ obtained by Bell measurements on two copies of the state at a time, without the need to access its conjugate. A key factor of both algorithms above is that, being limited in learning a discrete set of states obtained by the action of Clifford+T circuits, they learn an exact tomographic description in terms of Pauli observables. 

To enhance clarity, we summarize the results on the sample and computational complexity of both methods below. Since both algorithms have access to samples from a probability distribution on Pauli operators, $\Xi$ and $\tilde{\Xi}$ respectively, in what follows we split the sample complexity coming from sampling from such distributions from the sample complexity employed for additional Pauli measurements, which we refer to as \textit{measurement shots}. %However, while sampling from the characteristic distribution requires access to the queried state $\ket{\psi_t}$ and its conjugate in the computational basis $\ket{\psi_{t}^{*}}$, samples from $\tilde{\Xi}$ can be obtained by Bell measurements on two copies of $\ket{\psi_t}$ at a time.

\begin{algorithm} Assuming having access to sample from the characteristic distribution $\Xi_{\psi_t}$, the algorithm that learns a exact classical description of $t$-doped stabilizer state requires $O(2^tn+t2^{5t})$ $\Xi$-samples, $O(2^{7t}+2^{4t}n(n+t))$ measurement shots, $O(4^{2t}n^2)$ additional computational steps, and fails with probability $O(n2^{-n})$. 
\end{algorithm}

\begin{algorithm}
Having access to sample from $\tilde{\Xi}_{\psi_t}=d^{-1}|\!\braket{\psi_t|P|\psi_{t}^{*}}\!|^2$, the algorithm that learns a exact classical description of $t$-doped stabilizer state requires $O(2^{6t}n)$ $\tilde{\Xi}$-samples, $O(2^{9t}n(n+6t))$ measurement shots, $O(n^2)$ additional computational steps, and fails with probability $O(n2^{-n})$. 
\end{algorithm}

Furthermore, it is worth noting that for $t=0$, both algorithms reduce to Montanaro's algorithm.

%While the majority of our results hold for states doped by with $l$-local non-Clifford gate (with $l=O(1)$), we will focus our discussion on the $T$-gate case for clarity.

\section{Algebraic structure of $t$-doped stabilizer states}\label{sec:algstr}
Denote $\mathbb{P}_n$ the Pauli group on $n$ qubits and $\mathcal{C}_n$ the Clifford group. We define $t$-doped Clifford unitaries, denoted as $C_t$, unitary operators comprised of Clifford unitaries $C\in\mathcal{C}_n$ plus $t$ $l$-qubits non-Clifford gates with $l=O(1)$. Let $\ket{\psi_t}$ be a $t$-doped stabilizer state -- i.e. a state obtained as $\ket{\psi_t}=C_t\ket{0}^{\otimes n}$ -- and let $\psi_t$ be its density matrix. Define $S_{\psi_t}:=\{P\in\mathbb{P}_n\,|\, \tr(P\psi_t)\neq 0\}$ the set of Pauli operators with nonzero expectation over $\ket{\psi_t}$. For a stabilizer state ($t=0$) $|S_{\psi_{0}}|=d$. Further, one can define the probability distribution $\Xi_{\psi_t}$ with support on $S_{\psi_t}$ and components $\Xi_{\psi_t}(P)=\tr^2(P\psi_t)/d$. We denote the \textit{stabilizer entropy}~\cite{leone_stabilizer_2022} of the state $\psi_t$ as $M_{\alpha}(\psi_t)=S_{\alpha}(\Xi_{\psi_t})-n$,  that corresponds to the R\'enyi entropy $S_{\alpha}(\Xi_{\psi_t})$ of order $\alpha$ of the probability distribution $\Xi_{\psi}$ up to an offset $n$. As we shall see, the stabilizer entropy can be operationally interpreted as the entropy of the tomography in the Pauli basis.

Consider $M_0(\psi_t)\equiv\log(|S_{\psi_t}|/d)$. One has $M_{0}(\psi_t)\le 2lt$~\cite{leone_stabilizer_2022} and thus $|S_{\psi_t}|\le d2^{2lt}$. Furthermore, a $t$-doped stabilizer state $\psi_t$ admits a stabilizer group $G_{\psi_t}$~\cite{jiang_lower_2021}, i.e. a subgroup of $\mathbb{P}_n$ such that $\forall P\in G_{\psi_t}$  $P\ket{\psi_t}=\pm \ket{\psi_t}$.  However, unlike stabilizer states, the stabilizer group of a generic $t$-doped stabilizer state has a cardinality $|G_{\psi_t}|$ strictly less than $d$ and is bounded from below
\be
 \frac{d}{2^{2lt}}\le |G_{\psi_t}|\le d\,,
\ee
i.e. the set of Pauli operators stabilizing a given $t$-doped state gets (at most) halved every non-Clifford gate used to build $\ket{\psi_t}$ from $\ket{0}^{\otimes n}$. Since $G_{\psi_t}$ is a subgroup of the Pauli group, it is finitely generated and there exist $g_{1},\ldots, g_{m}$ generators, with $m\equiv \log_2|G_{\psi_t}|$, such that every element $Q\in G_{\psi_t}$ can be obtained by a product of $g_{1},\ldots, g_{m}$ with relative phases $\phi_{g_1},\ldots, \phi_{g_m}$ defined by $g_{i}\ket{\psi_t}=\phi_{g_i}\ket{\psi_t}$. Since there are at most $|S_{\psi_t}|\le d2^{2lt}$ Pauli operators with nonzero expectation over $\psi_t$ and at least $d/2^{2lt}$ of those stabilize $\ket{\psi_t}$, there exist Pauli operators $h_{1},\ldots, h_{k}\in S_{\psi_t}$ ( $k\le 4^{2lt}-1$) such that the set $S_{\psi_t}$ can be written as the union of left disjoint cosets (see Appendix~\ref{app:proofs_dc})
\be\label{structureofS}
S_{\psi_t}=G_{\psi_t}\cup h_{1}G_{\psi_t}\cup\ldots\cup h_{k}G_{\psi_t}\,.
\ee
Throughout the manuscript, we refer to the coset representatives $h_1,\ldots, h_k$ as \textit{bad-generators} of the $t$-doped stabilizer state $\psi_t$. Note that the number $m$ of generators $g_i$ and the number $k$ of bad-generators $h_i$ must obey  $2^m(k+1)=|S_{\psi_t}|$. Thanks to the structure of $S_{\psi_t}$ in Eq.~\eqref{structureofS}, a $t$-doped stabilizer state can be expressed in the Pauli basis as
\be
\psi_t=\frac{1}{d}\sum_{i=0}^{k}\tr(h_{i}\psi_t)h_i\prod_{j=1}^{m}\left(\id+\phi_{g_j}g_{j}\right)\,,
\label{pgphy}
\ee
where $h_{0}\equiv \id$. Notice that the operator $\prod_{j=1}^{m}(\id+\phi_{g_j}g_j)=2^m \Pi_{\psi_t}$ is proportional to the projector $\Pi_{\psi_t}$ onto the stabilizer group $G_{\psi_t}$, which automatically implies that bad generators commute with every $P\in G_{\psi_t}$ to ensure hermiticity and positivity of $\psi_t$. Through the decomposition in Eq.~\eqref{pgphy}, we can easily derive the corollary of the results in~\cite{leone_learning_2022,oliviero_black_2022} that will be useful later on. Define $\mathcal{D}\in \mathcal{C}_n$ a Clifford operator that obeys $\mathcal{D}\Pi_{\psi_t}\mathcal{D}^{\dag}=\st{0}^{\otimes m}\otimes \id_{[n-m]}$, where $\id_{[n-m]}$ denotes the identity on the last $n-m$ qubits. Then since $h_i$ are coset representative and do commute with $\Pi_{\psi_t}$, the following decomposition holds
\be
\ket{\psi_t}=\mathcal{D}^{\dag}(\ket{0}^{\otimes m}\otimes \ket{\phi})\,,\label{diagonalizermain}
\ee
where $\ket{\phi}$ is a state defined on the last $(n-m)$ qubits. The density matrix associated to $\ket{\phi}$ can be written as $\st{\phi}=2^{n-m}\sum_{i=0}^{k}\tr(h_i\psi_t)\tilde{h}_i$, where $\tilde{h}_i\equiv\mathcal{D}h_i\mathcal{D}^{\dag}$. Moreover the following relation, useful for what follows, holds $G_{\mathcal{D}\psi_t\mathcal{D}^{\dag}}=\mathcal{D}G_{\psi_t}\mathcal{D}^{\dag}$. To be consistent with the terminology used in Refs.~\cite{leone_learning_2022,oliviero_black_2022}, we refer to the Clifford operator $\mathcal{D}$ as the \textit{diagonalizer}.

As another consequence of Eq.~\eqref{pgphy}, a direct computation of the stabilizer entropy $M_{\alpha}(\psi_t)$ returns $M_{\alpha}(\psi_t)=E_{\alpha}(\chi_{\psi_t})-\nu(\psi_t)$, where $\nu(\psi_t)\equiv n-m$ is the stabilizer nullity introduced in~\cite{beverland_lower_2020}, while $E_{\alpha}(\chi_{\psi_t})$ is the $\alpha$-R\'enyi entropy of the probability distribution $\chi_{\psi_t}$ with support on the cosets $h_iG_{\psi_t}$ and components $\chi_{\psi_t}(h_i)=\tr^2(h_i\psi_t)|G_{\psi_t}|/d$. The probability distribution $\chi_{\psi_t}$ can be interpreted as the entropy of bad generators $h_{1},\ldots, h_{k}$, e.g. $E_{0}(\psi_t)=\log (k+1)$. 

While all the results presented so far are general, let us conclude the section by directing our attention toward $t$-doped stabilizer states resulting from the application of Clifford+T circuits.  For T-gates, the previously derived bounds can be obtained by substituting $2l$ with $1$. Indeed, $l=1$ for single-qubit gates, and the additional $1/2$ factor arises when considering diagonal gates, see~\cite{jiang_lower_2021}. As these states form a discrete set, it is expected that the expectation values of Pauli operators exhibit discrete values. Hence, there should exist a finite resolution, strictly dependent on $t$, enabling one to exactly distinguish these values by employing a sufficient number of samples. This intuition is made rigorous in the following discussion. Consider a $t$-doped stabilizer state obtained by the action of $t$ $T$-gates. Then, for every $P,Q\in\mathbb{P}_n$, the following lower bound holds
\be
|\tr(P\psi_t)-\tr(Q\psi_t)|\ge\frac{\sqrt{2}}{6}\left(\frac{1}{\sqrt{2}}-\frac{1}{2}\right)^{t}
\label{eq:finiteresolution}
\ee
provided that $\tr(P\psi_t)\neq\tr(Q\psi_t)$. See Appendix~\ref{proof_fr} for the proof. Therefore, for $t=O(\log n)$, the expectation values of Pauli operators either coincide or exhibit a polynomial separation. This fact will enable us to say that by employing a sufficient number of samples -- though polynomial in $n$ for $t=O(\log n)$ --  to measure Pauli expectation values, one can learn $t$-doped stabilizer states \textit{exactly}. Let us remark again that, contrary to all statements presented in the section, Eq.~\eqref{eq:finiteresolution} does not hold for the broader scenario of doping with $O(1)$-qubit non-Clifford gates; rather, it specifically applies to the case involving $T$-gates exclusively.

\section{Learning algorithm from $\Xi_{\psi_t}$-samples}\label{sec:learning1}
In the previous section, we explicitly revealed the algebraic structure of a generic $t$-doped stabilizer state. Let us provide the algorithm able to learn a $t$-doped stabilizer state. While the majority of our results can be generalized to states doped by $l$-local non-Clifford gates (with $l=O(1)$), we will focus our discussion on the $T$-gate case in order to employ the finite resolution result in Eq.~\eqref{eq:finiteresolution}. In particular, a state $\psi_t$ doped with a number $t$ of $T$-gates is characterized by $(i)$ $m\ge n-t$ generators $g_{i}$ of $G_{\psi_t}$ with $m$ relative phases $\phi_{g_i}$; $(ii)$ by $k\le 4^{t}-1$ bad generators $h_i$ with $k$ expectation values $\tr(h_i\psi_t)$. Therefore, the learning of a classical description of a $t$-doped stabilizer state requires the knowledge of $2m+2k\le 2n+2\times 4^t$ objects, and therefore for $t=O(\log n)$ we have $2m+2k=O(\poly(n))$\footnote{Essentially, the complexity stems from the fact that the support $|S_{\psi_t}|\le 2^{t}d$. However, it's worth noting that a weaker form of such a bound, i.e., $|S_{\psi_t}|\le 3^{t}d$, was previously demonstrated in Ref.~\cite{lai_learning_2022}.}.

Let us assume to be able to sample from the distribution $\Xi_{\psi_t}$. Later on, in Section~\ref{sec:sampling}, we discuss in which situations an efficient sampling from $\Xi_{\psi_t}$ is possible by querying the $t$-doped state $\psi_t$. We first aim to learn a basis for the stabilizer group $G_{\psi_t}$, i.e. $m$ generators $g_{i}$. Sampling from $\Xi_{\psi_t}$ we get a Pauli operator $P\in G_{\psi_t}$ with probability
\be
\operatorname{Pr}(P\in G_{\psi_t})=\frac{|G_{\psi_t}|}{d}\ge\frac{1}{2^t}\,.\label{eq.2}
\ee
Therefore, after $O(2^t)$ samples one gets a Pauli operator sampled uniformly from $G_{\psi_t}$ with high probability. To check whether a sampled Pauli operator $P$ stabilizes $\ket{\psi_t}$ is sufficient to measure $M$ times $P$. If all the measurement outcomes are equal, then $P\in G_{\psi_t}$ with failure probability at most $(1-2^{-t})(h_{\max}/2+1/2)^{M}$ (see Appendix~\ref{app:proofs}), where $h_{\max}\equiv\max_{h_i}|\tr(h_i\psi_t)|$. We remark that the above procedure also reveals the stabilizing phase $\phi_{P}$ corresponding to $P\in G_{\psi_t}$. Following the reasoning introduced by Montanaro~\cite{montanaro_learning_2017}, the extraction of $m+n$ random Pauli operators  $\in G_{\psi_t}$ suffices to identify a set of generators $g_{1},\ldots, g_{m}$ with failure probability at most $2^{-n}$. Indeed, determining a basis for $G_{\psi_t}$ is equivalent to determining a basis for the $m$-dimensional subspace $T$ of $\mathbb{F}_{2}^{2n}$; the probability that $n+m$ random samples from $T$ lie in a $(m-1)$-dimensional subspace of $T$ is $2^{-m-n}$ and -- by the union bound -- the probability that these samples lie in one of the $2^{m}$ $(m-1)$-dimensional subspaces of $T$ is at most $2^{-n}$. We conclude that $O(2^t(n+m))$ samples from the probability distribution $\Xi_{\psi_t}$, plus $O(2^{t}(n+m)M)$ additional measurement shots, are sufficient to learn all the generators $g_{1},\ldots, g_{m}$ and phases $\phi_{g_i}$ with failure probability $O(2^t(n+m)(h_{\max}/2+1/2)^{M}+2^{-n})$. 

At this point, let us learn the $k$ bad generators $h_1,\ldots, h_{k}$. Suppose the algorithm already found $l$ out of $k$ bad generators, say $h_{1},\ldots, h_{l}$. To find the $(l+1)$-th bad generator, we keep sampling from $\Xi_{\psi_t}$. The probability $\pi_l$ that the sampled Pauli operator $P$ does not belong to $G_{\psi_t}\cup h_{1}G_{\psi_t}\cup \ldots\cup h_{l}G_{\psi_t}$ is 
\be\label{probabilitypil2}
\pi_l=1-\frac{|G_{\psi_t}|}{d}\left(1+\sum_{i=1}^{l}\tr^2(h_{i}\psi_t)\right)\,.
\ee
To find a useful lower bound to $\pi_l$, let us exploit the unit purity of the state $\psi_t$. In Appendix~\ref{app:proofs}, we obtain
\be
\pi_l\ge |G_{\psi_t}|(k-l)h_{\min}^2/d\label{probabilitypil}
\ee
with $h_{\min}:=\min_{h_i}|\tr(h_i\psi_t)|$.
However, a learner needs an efficient way to check whether a sampled Pauli operator belongs to  $G_{\psi_t}\cup h_{1}G_{\psi_t}\cup \ldots\cup h_{l}G_{\psi_t}$.
Let us provide a simple algorithm to efficiently check whether a Pauli operator $P$ belongs to $G_{\psi_t}\cup h_{1}G_{\psi_t}\cup \ldots\cup h_{l}G_{\psi_t}$ or not. 
Let us assume $P\in h_0G_{\psi_t}\cup h_1G_{\psi_t}\cup\cdot\cup h_{l}G_{\psi_t}$, where $h_0=\id$. Then for some $i\in 0\ldots l$, there exists $h_i$ such that $ h_iP\in G_{\psi_t}$. Thus the problem reduces to check if there exists an $h_i\in\{h_0,\ldots,h_l\}$ such that $h_iP\in G_{\psi_t}$. The latter can be solved by adding $h_iP$ to a generating set $G_{\psi_t}$ and performing Gaussian elimination over $\mathbb{F}_2^{2n}$. The task requires $O(n^3)$ steps and has to be repeated $l$ times, meaning that the full task requires $O(n^3l)$ computational steps, and so to learn all the $k$ bad generators one needs $O(n^3k^2)$. In appendix~\ref{app:cont} we provide a slightly more efficient algorithm relying instead on the notion of diagonalizer\cite{leone_learning_2022,oliviero_black_2022} where the number of computational steps to verify the containment of one bad generator is $O(n^2k)$, and so reducing the computational steps to learn all the bad generators to $O(n^2k^2)$. The total number of samples scales as $\sum_{l=0}^{k-1}\pi_{l}^{-1}\le 2^{t}h_{\min}^{-2}\sum_{l=0}^{k-1}(k-l)^{-1}\le 2^th_{\min}^{-2}(\gamma+\log(k+1))$ for $\gamma$ being the Euler's constant. Therefore using $O\left(2^{t}h_{\min}^{-2}\log(k+1)\right)$ samples from $\Xi_{\psi_t}$, plus $O(n^2k^2)$ computational steps, one learns the $k$ bad generators $h_{1},\ldots, h_{k}$. To evaluate $h_{\min}$ and $h_{\max}$, let us recall that the state $\ket{\psi_t}$ is obtained from the computational basis state $\ket{0}^{\otimes n}$ by a Clifford circuit polluted with $t$ single qubit $T$-gates, that generate a discrete set of states, which ultimately implies the existence of a finite resolution $\delta_t\equiv\min|\tr[(P-Q)\psi_t]|$ for $\tr(P\psi_t)\neq\tr(Q\psi_t)$. From Eq.~\eqref{eq:finiteresolution}, one has $\delta_t=\Omega(2^{-bt})$ for $b\simeq 2.27$. Therefore $O(2^{2bt})$ measurement shots are sufficient to determine the expectation value $\tr(h_{i}\psi_t)$ exactly. In addition, one has $h_{\min}=\Omega(2^{-bt})$ and $h_{\max}<1-O(2^{-bt})$. In summary, to learn $k$ bad generators and the corresponding expectations, the learner employs $O(2^{(2b+1)t}\log(k+1))$ samples from $\Xi_{\psi_t}$, $O(k2^{2bt})$ measurement shots, and $O(n^2k^2)$ additional computational steps.

Let us count the total number of resources to learn a $t$-doped stabilizer state employing the algorithm proposed above. Set $m\le n$, $k< 4^t$, $b<3$ and $M=2^{3t+1}(n+t)$. The total number of samples from the probability distribution $\Xi_{\psi_t}$ is $O(2^{t}n+t2^{5t})$ plus $O(4^tn^2)$ additional computational steps. The total number of measurement shots is $O(2^{7t}+2^{4t}n(n+t))$ and the algorithm fails with probability at most $O(n2^{-n})$. Therefore for $t=O(\log n)$, the algorithm learns the tomographic description of a $t$-doped stabilizer state of Eq.~\eqref{pgphy} with polynomial resources and overwhelming probability.

\section{Sampling from $\Xi_{\psi_t}$}\label{sec:sampling}
In Section~\ref{sec:learning1}, we provided an algorithm that learns a $t$-doped stabilizer state with $\operatorname{poly}(n,2^t)$ resources. In particular, the algorithm uses $O(2^tn+t2^{5t})$ samples from the probability distribution $\Xi_{\psi_t}$ with elements $\Xi_{\psi_t}(P)=\tr^2(P\psi_t)/d$. Let us discuss the connection between queries to the unknown state $\ket{\psi_t}$ and samples from the distribution $\Xi_{\psi_t}$. Having query access to the $t$-doped stabilizer state $\ket{\psi_t}$ and its conjugate (in the computational basis) $\ket{\psi_{t}^{*}}$, one can easily achieve the task via Bell sampling. Define the Bell basis on two copies of Hilbert space $\mathcal{H}$ as $\ket{P}\equiv \id\otimes P\ket{\id}$ with $\ket{\id}=2^{-n/2}\sum_{i=1}^{2^n}\ket{i}\otimes \ket{i}$ and $P\in\mathbb{P}_n$. Then, measuring $\ket{\psi_t\otimes \psi_{t}^*}$ in the basis $\ket{P}$ is equivalent to sample from the distribution $\Xi_{\psi_t}$:
\be
\operatorname{Pr}(\ket{\psi_t\otimes\psi_{t}^*}\mapsto \ket{P})=\Xi_{\psi_t}(P)\,.
\ee
Therefore a single query to $\ket{\psi_t}$ and its conjugate $\ket{\psi_{t}^{*}}$ suffices to obtain one sample from the distribution $\Xi_{\psi_t}$. 

At this point, the careful reader may wonder how the learner can prepare the conjugate state $\ket{\psi_{t}^{*}}$ in order to use the algorithm. We have a manifold of answers to this. First of all, an important situation in which learning is important is that of the study of ground states of quantum many-body systems. An example is provided by stabilizer Hamiltonians~\cite{coble_hamiltonians_2023} perturbed by local impurities, which exhibit $t$-doped stabilizer eigenstates as shown in~\cite{gu2024doped}. Generically, as long as time-reversal is not broken, these are real states, and thus $\ket{\psi_{t}^{*}}=\ket{\psi_{t}}$. Moreover, in many scenarios, the preparation itself of the state $\ket{\psi_{t}}$ can be thought of as made by a quantum circuit. In this case then, the task of preparing $\ket{\psi_{t}^{*}}$ is straightforward. Indeed, it is sufficient to replace the gate $S=\operatorname{diag}(1,i)$ with $S^{*}=\operatorname{diag}(1,-i)$ and the $T$-gate with $TS^{*}$ to obtain $U_{t}^{*}$ and thus constructing $\ket{\psi_{t}^{*}}$ from $\ket{0}^{\otimes n}$. 

In any case, while distilling the conjugate state is a hard task in general, the states under examination in this paper for which the learning is efficient, i.e. $t$-doped stabilizer states with $t=O(\log n)$, have fine structure and hardness proofs fail at capturing the hardness of distilling $\ket{\psi_{t}^{*}}$ from samples of $\ket{\psi_t}$. For example, the hardness proof in Ref.~\cite{haug_pseudorandom_2023} uses pseudorandom quantum states and it is known that pseudorandom quantum states contain $\omega(\log n)$ many $T$-gates~\cite{grewal2024improved}. Indeed, from the knowledge of the algebraic structure in Eq.~\eqref{pgphy}, sampling from the characteristic distribution is straightforward: it is sufficient to flip a $k$-faceted dice with probabilities $\chi_{\psi_t}(h_i)$ -- defined above in Section~\ref{sec:algstr} -- and outcomes $h_i$, and uniformly sample a Pauli operator $P\in G_{\psi_t}$ through $m$ generators. As a result, one samples a Pauli operator $h_iP$ according to $\Xi_{\psi_t}$. This procedure is readily efficient for $k=O(\poly(n))$, i.e. for $t=O(\log(n))$. Therefore, for $t=O(\log n)$, there is no reason to believe that sampling from the distribution $\Xi_{\psi_t}$, with queries limited to $\ket{\psi_t}$ alone and without knowing the algebraic structure~\eqref{pgphy} (i.e. bypassing the a priori learning of the $t$-doped stabilizer state), is hard and, as such, it remains an ongoing subject of research. Remarkably, in Ref.~\cite{hinsche2024efficient}, an algorithm is presented capable of sampling from the characteristic distribution that includes, but is not limited to, states with $t=O(\log n)$ and low entanglement.

\section{Learning algorithm without access to $\Xi_{\psi_t}$}\label{sec:learning2}
We discussed a simple and intuitive learning algorithm that works whenever one has access to samples from the distribution $\Xi_{\psi_t}$. However, as we saw in Section~\ref{sec:sampling}, known methods for sampling from $\Xi_{\psi_t}$ require query access to the $t$-doped stabilizer state and its conjugate in the computational basis, which is not always available in real scenarios. In this section, we present an algorithm similar in spirit to the one in Section~\ref{sec:learning1}, but we relax the hypothesis of having samples from $\Xi_{\psi_t}$.

As measuring $\ket{\psi_t\otimes\psi_t^{*}}$ in the Bell basis $\ket{P}$ returns samples from $\Xi_{\psi_t}$, measuring $\ket{\psi_t\otimes \psi_t}$ in the Bell basis returns samples from 
\be
\widetilde{\Xi}_{\psi_t}(P)\equiv\frac{|\braket{\psi_t|P|\psi_t^{*}}|^2}{d}\,.\label{probabilityreal}
\ee
Let us show that samples from $\widetilde{\Xi}_{\psi_t}$ are still sufficient to learn the stabilizer group $G_{\psi_t}$ and, ultimately, the $t$-doped stabilizer state $\psi_t$ exactly. To see this, it is useful to show how the algebraic structure changes from $\psi_t$ to $\psi_{t}^{*}$. Naively, from Eq.~\eqref{pgphy}, we can write the density matrix associated to $\psi_{t}^{*}$ as
\be
\psi_{t}^{*}=\frac{1}{d}\sum_{i=0}^{k}\tr(h_i\psi_t^*)h_i\prod_{j=1}^{m}(\id+\phi_{g_j}g_{j}^{*})\,.
\ee
To understand the algebraic structure of $\psi_{t}^{*}$, it is useful to look back at the simple case of stabilizer states, i.e. $t=0$. As shown by Montanaro in Ref.~\cite{montanaro_learning_2017}, given a stabilizer state $\ket{\sigma}$, its conjugate $\ket{\sigma^*}$ can be obtained with a Pauli rotation as $\ket{\sigma^{*}}=P_{\sigma}\ket{\sigma}$ for some Pauli operator $P_{\sigma}$. Similarly, since the projector $\Pi_{\psi_t}$ in Eq.~\eqref{pgphy}, onto the stabilizer group $G_{\psi_t}$, can be completed to represent a pure stabilizer state (by completion of the stabilizer group $G_{\psi_t}$), it is immediate to see that there exist a Pauli operator $P_{\psi_t}$ such that (see Appendix~\ref{App:sampling}) 
\be
P_{\psi_t}\Pi_{\psi_t}P_{\psi_t}=\Pi_{\psi_t}^{*}\,,
\ee
where $\Pi_{\psi_t}^{*}\equiv 2^{-m}\prod_{j=1}^{m}(\id+\phi_{g_j}g_{j}^{*})$. In other words, analogously to the case of stabilizer states, the projectors onto the stabilizer groups $\Pi_{\psi_t},\Pi_{\psi_t}^{*}$, of $\psi_t$ and $\psi_{t}^{*}$ respectively, can be obtained from one another by the application of a Pauli operator $P_{\psi_t}$. Hence, the state $\ket{\psi_{t}^{*\prime}}:=P_{\psi_t}\ket{\psi_{t}^{*}}$ has the same stabilizer group $G_{\psi_t}$ of $\psi_t$, by construction.

As explained in Section~\ref{sec:algstr}, from the projector $\Pi_{\psi_t}$ onto the stabilizer group $G_{\psi_t}$, one can define a diagonalizer $\mathcal{D}$. Since the stabilizer groups are the same, $\mathcal{D}$ diagonalizes $\ket{\psi_{t}}$ and $\ket{\psi_{t}^{*\prime}}$ at the same time as $\mathcal{D}\ket{\psi_t}=\ket{0}^{\otimes m}\otimes\ket{\phi}$ and $\mathcal{D}\ket{\psi_{t}^{*\prime}}=\ket{0}^{\otimes m}\otimes\ket{\phi^{*}}$, cfr. Eq.~\eqref{diagonalizermain}. By simple manipulations, we can thus rewrite the probability in Eq.~\eqref{probabilityreal} as only depending on the nonstabilizer part $\ket{\phi}$ of the $t$-doped stabilizer state $\ket{\psi_t}$ as
\be
\widetilde{\Xi}_{\psi_t}(P)=\frac{|\braket{\phi|\bar{P}|\phi^{*}}|^2}{d}\,,\label{prob2}
\ee
with $\bar{P}\equiv\braket{0^{\otimes m}|\mathcal{D}PP_{\psi_t}\mathcal{D}^{\dag}|0^{\otimes m}}$. It is easy to be convinced that the probability $\tilde{\Xi}_{\psi_t}(P)$ is different from zero only if $PP_{\psi_t}\in \mathcal{D}^{\dag}(\id_m\otimes \bar{P})\mathcal{D}G_{\psi_t}$ for any $\bar{P}\in\mathbb{P}_{n-m}$ living on the last $(n-m)$ qubits. Hence, samples from $\widetilde{\Xi}_{\psi_t}(P)$ returns Pauli operators $P\in\{\mathcal{D}^{\dag}(\id_{m}\otimes\bar{P})\mathcal{D}G_{\psi_t}\,|\,\bar{P}\in\mathbb{P}_{n-m}\}$ with probability \eqref{prob2}. As in the case of stabilizer states, to gain knowledge of the stabilizer group $G_{\psi_t}$, it is sufficient to obtain two Pauli operators $P,P^{\prime}$ belonging to the same left coset $P,P^{\prime}\in \mathcal{D}^{\dag}(\id_{m}\otimes\bar{P})\mathcal{D}G_{\psi_t}$, and consider the product $PP^{\prime}\in G_{\psi_t}$. At this point, the joint probability that two samples give a product Pauli operator belonging to $G_{\psi_t}$ is lower bounded by
\be
\operatorname{Pr}(PP^{\prime}\in G_{\psi_t})\ge\frac{1}{2^{6t}}
\ee
where the proof is in Appendix~\ref{App:sampling}.
Therefore, repeating the reasoning outlined in Section~\ref{sec:learning1} after Eq.~\eqref{eq.2}, we conclude that $O(2^{6t}(n+m))$ samples from $\widetilde{\Xi}_{\psi_t}$, plus $O(2^{9t}n(n+6t))$ additional measurement shots, are sufficient to learn all the generators $g_{1},\ldots, g_{m}$ and phases $\phi_{g_{i}}$ with exponentially small probability of failure $O(n2^{-n})$. 

Having learned the generators $g_{i}$ and relative phases $\phi_{g_i}$, it is thus possible, through standard tableau manipulation techniques (see Ref.~\cite{leone_learning_2022}), to construct and distill the diagonalizer $\mathcal{D}$ in Eq.~\eqref{diagonalizermain} using $O(n^2)$ computational steps and up to $O(n)$ Clifford gates~\cite{bravyi_hadamardfree_2021}. Applying $\mathcal{D}$ on the $t$-doped stabilizer state $\ket{\psi_t}$, one gets $\mathcal{D}\ket{\psi_t}=\ket{0}^{\otimes m}\otimes\ket{\phi}$. At this point, one is left to learn the tomographic description of $\ket{\phi}$ in terms of Pauli operators $\tilde{h}_i\equiv\mathcal{D}h_i\mathcal{D}^{\dag}$. To achieve this task, it is sufficient to individually measure each Pauli operator $\tilde{h}_i$ for $i=1,\ldots, k$. In virtue of Eq.~\eqref{eq:finiteresolution}, to estimate the expectation values $\tr(h_i\psi_t)$ exactly, $O(2^{2bt})$ samples are sufficient, and thus to learn the expectation values of $k$ bad generators $h_i,\ldots, h_k$ one needs $O(4^{t}2^{2bt})$  total measurement shots. In summary, the above algorithm learns an exact tomographic description of a $t$-doped stabilizer state using $O(2^{6t}n)$ samples from the distribution $\tilde{\Xi}$ (which in turn can be obtained by measuring $2$ copies of $\psi_t$ in the basis $\ket{P}$), $O(2^{9t}n(n+6t))$ measurement shots, $O(n^2)$ additional computational steps, and fails with probability $O(n2^{-n})$. 

The careful reader might have noticed similarities between the procedure outlined above and the subroutine known as \textit{Bell difference sampling}~\cite{gross_schur_2021}, where one considers two samples from the distribution $\widetilde{\Xi}_{\psi_t}$ and, subtracting the results, obtain sample Pauli operators $P\in G_{\psi_t}^{\perp}=\{P\in\mathbb{P}\,|\, [P,G_{\psi_t}]=0\}$ whose knowledge allows to reconstruct the stabilizer group $G_{\psi_t}$, as shown in~\cite{grewal_efficient_2023}. There are two key insights here: first, the derivation outlined above solely descends from the algebraic structure of $t$-doped stabilizer states developed in this paper that provide a simple proof of the results. Secondly, the algorithm outlined above and the Bell difference sampling routine exhibit similarities, yet they are not entirely identical. Specifically, the above algorithm exclusively accommodates pairs of Pauli operators, denoted as $P$ and $P^{\prime}$, whose product resides within $G_{\psi_t}\subset G_{\psi_{t}}^{\perp}$. Hence, it does not need any classical postprocessing. This fact allows for the complete and precise determination of the stabilizer group $G_{\psi_t}$ with overwhelming probability, albeit at the expense of an exponential increase in computational cost with respect to $t$, a scaling that Bell difference sampling does not possess. %Nevertheless, the above algorithm indeed draws inspiration from the broader and more universally applicable routine of Bell difference sampling.

\section{Concurrent work}
During the finalization of this manuscript, we became aware of two works~\cite{grewal_efficient_2023,hangleiter_bell_2023} sharing several similarities with ours, presenting an efficient approach for learning $t$-doped stabilizer states. This section aims to compare algorithms. The algorithms presented separately in \cite{grewal_efficient_2023} and \cite{hangleiter_bell_2023} are quite similar. Hence, we focus on comparing the one from \cite{grewal_efficient_2023} with ours for simplicity.

The algorithm in \cite{grewal_efficient_2023} has a significant advantage: it uses Bell difference sampling to approximate the stabilizer group $G_{\psi_t}$ associated with the $t$-doped stabilizer state $\ket{\psi_t}$ using $O(n)$ resources, avoiding exponential scaling with the number $t$ of non-Clifford gates. However, it only learns an approximate version of the stabilizer group, which might not suit cases where an exact description of a $t$-doped stabilizer state is required. For T-gates, exact learning is desirable since these states are countable and form a discrete set. Whether the algorithm~\cite{grewal_efficient_2023} can be employed to learn the exact stabilizer group in light of the finite resolution proven in Eq.~\eqref{eq:finiteresolution} is not clear. The algorithm presented in Section~\ref{sec:learning2} of this paper, leveraging the algebraic structure introduced here, can precisely learn the stabilizer group $G_{\psi_t}$ from copies of $\ket{\psi_t}$, with an exponentially small probability of failure. However, this precise learning comes at the cost of exponential resource scaling with the number of T-gates. Compared to the algorithm in \cite{grewal_efficient_2023}, our algorithm's primary strength lies in its exact learning ability for states created by Clifford+T circuits, enabled by Eq.~\eqref{eq:finiteresolution}. Both algorithms in Section~\ref{sec:learning1} and~\ref{sec:learning2}, with different methods, whose limitations have been explored above, achieve the exact learning of the algebraic structure of $t$-doped stabilizer state displayed in Eq.~\eqref{pgphy}.  However, it's important to note that these states are not the only ones with polynomially many bad generators, and therefore our algorithms do not cover the larger class of states learnable according to the algorithm in Ref.~\cite{grewal_efficient_2023}, which, in this regard, is more general. Overall, we conclude that the two algorithms exhibit distinct ranges of applicability and employ different techniques to achieve the same task, each with its own set of strengths and limitations.

\section{Conclusions}
In conclusion, we presented an efficient and general algorithm that can learn a classical description of states obtained from the computational basis by the action of the Clifford group plus a few $T$ gates. To address this challenge, we introduced a new algebraic structure for $t$-doped stabilizer states that is of independent interest and plays a crucial role in the learning algorithm. 

In perspective, there are several open questions. First, we ask whether the algebraic structure presented in this paper can be instrumental in solving tasks beyond quantum state tomography for $t$-doped stabilizer state, e.g. in computing quantity as entanglement or magic. Second, we would like to know whether the knowledge of both the generators and the bad generators can be utilized to synthesize a Clifford+$T$ circuit able to construct the state $\ket{\psi_t}$ from $\ket{0}^{\otimes n}$. Indeed, note that the task is conceivable for two reasons: 1) because the complexity (i.e. the number of gates) of Clifford$+T$ circuits scales as $O(n^2+t^3)$~\footnote{By Theorem $2$ of~\cite{leone_learning_2022} a Clifford+$T$ circuit $C_t$ with $t\le n$ has the following decomposition $C_t=C_1(\id\otimes u_t) C_2$, with $C_1,C_2\in \mathcal{C}_n$ and $u_t$ a $t$-doped Clifford circuit on $t$-qubits consisting of at most $t$ many $T$ gates. From~\cite{bravyi_hadamardfree_2021}, any Clifford circuit can be implemented with at most $O(n^2)$ gates. Thus, $\#(C_i)= O(n^2)$ for $i,1,2$ and $\#(u_t) =O(t^3)$, and consequently the total complexity is $O(n^2+t^3)$ } as a consequence of the results of~\cite{oliviero_black_2022,leone_learning_2022,bravyi_hadamardfree_2021}; 2) given a $t$-doped stabilizer state, one can learn exactly the Clifford circuit $\mathcal{D}$, reducing the task of learning the circuit description to $u_t$ only. Lastly, the estimated resources are computed for the worst case. We ask whether the average case complexity for the exact learning of $t$-doped stabilizer states doped with $T$-gates is more favorable in terms of the exponential scaling in $t$.

%\section*{Note added} During the finalization of this manuscript, we became aware of two works~\cite{grewal_efficient_2023,hangleiter_bell_2023} sharing several similarities with ours, presenting an efficient approach for learning $t$-doped stabilizer states. However, while our approach uses different methods than the other two works, each approach has its own distinct strengths and limitations.

\section*{Acknowledgments} The authors acknowledge discussions with M. Hinsche, M. Ioannou, S. Jerbi, J. Carrasco and G. Esposito. The authors would like to express their gratitude to Daniel Miller for his valuable insight regarding the existence of a finite resolution for t-doped stabilizer states. The authors thank the anonymous Reviewer $1$ for their constructive comments and suggestions that helped to improve the manuscript. Special thanks go to Lennart Bittel for fundamental inceptions for the proof of the lower-bound in Eq.~\eqref{eq:finiteresolution}. AH was supported by the ENEA project I53C22003030001, the PNRR MUR project PE0000023-NQSTI and PNRR MUR project CN $00000013$-ICSC.  L.L. S.F.E.O. acknowledge support from NSF award number 2014000. L.L. and S.F.E.O. contributed equally to this work.
\bibliographystyle{quantum}
\bibliography{bib}

\appendix
\onecolumngrid
\section{Technical preliminaries}
In this section, we will introduce some technical preliminaries that may prove beneficial to the reader. 
\subsection{Pauli Group, Clifford group, and stabilizer states}
The Pauli group $\mathcal{P}_n$ is the $n$-tensor fold product of the single-qubit Pauli group $\mathcal{P}_1$, which consists of the elements ${\id, X, Y, Z}$, each multiplied by a scaling factor of ${\pm 1, \pm i}$, where $X,Y$ and $Z$ are the standard single-qubit Pauli matrices. Let us introduce the quotient group $\mathbb{P}_n$ of the Pauli group $\mathbb{P}:=\mathcal{P}/\{\pm 1 ,\pm i \}\,$. In the main text, we refer to the quotient group as Pauli group.
The Clifford group $\mathcal{C}(n)$ on $n$ qubit, i.e. a subgroup of the unitary group is defined as the normalizer of the Pauli group $\mathcal{P}_n$:
\be
\mathcal{C}_n:=\{ C\in\mathcal{U}(n)\,\,\, |\,\,\, C^{\dag}PC\in\mathcal{P}, \quad \forall P\in\mathcal{P}_n\}.
\ee
With the notion of Pauli and Clifford group, one can then define the notion of a stabilizer state. Given a Pauli operator $P\in\mathbb{P}_n$, we say that $P$ stabilizes the state $\ket{\psi}$ if 
\be 
P\ket{\psi}=\pm \ket{\psi}
\ee 
A state $\ket{\sigma}$ is a stabilizer states if $\ket{\sigma}$ is stabilized by an group $G$ of $d$ commuting Pauli operators
\be 
P\ket{\sigma}=\pm\ket{\sigma}\quad \forall P\in G
\ee 
Equivalently, one can define a stabilizer state as $\ket{\sigma}\equiv C\ket{0}^{\otimes n}$ with $C\in\mathcal{C}_n$. It is noteworthy to notice that a stabilizer state can be written as an equal superposition of Pauli operators that stabilizes it 
\be
\sigma =\frac{1}{d}\sum_{P\in G}\sigma_P P
\ee 
where $\sigma$ is the density matrix associated to $\ket{\sigma}$ and $\sigma_p=\pm 1$ is the phase associated to $P$. Being $G$ an Abelian group, there exists a set of generators $S\equiv\{g_1,\ldots,g_n\}$ such that $G=\langle S\rangle$. Thus, given $S$ the set of generators associate with $G$ the state $\sigma$ can be expressed in terms of the generator $g_i\in S$ as
\be 
\sigma = \prod_{i}\left(\frac{\id+\phi_{g_i}g_i}{2}\right)
\ee 
where $\phi_{g_i}$ are the $\pm 1$ phases associated with the generators $g_i$s. 

\subsection{Bell Sampling}
In this section, we will review the concept of Bell sampling~\cite{montanaro_learning_2017} and its application in learning a stabilizer state. To grasp the idea behind Bell sampling, let's begin by considering the $2$-qubit maximally entangled state, denoted as $\ket{\id}$, which is defined as $(\ket{00}+\ket{11})/\sqrt{2}$. If we apply the operator $\id\otimes P$, where $P={\id,X,Y,Z}$, to the state $\ket{\id}$, we obtain the set of states: $\ket{\id}$, $\ket{X}=(\ket{01}+\ket{10})/\sqrt{2}$, $\ket{Y}=(\ket{01}-\ket{10})/\sqrt{2}$, and $\ket{Z}=(\ket{00}-\ket{11})/\sqrt{2}$. This set of states corresponds to the $2$-qubit Bell basis. This equivalence between Pauli operators and states can be extended to the $n$-qubit case. The reason behind this extension lies in the fact that the $n$-qubit Pauli group is constructed as the $n$-fold tensor product of the $1$-qubit Pauli group. Consequently, we can tensor $2$-qubit states $\ket{P}$ to establish an equivalence between the $n$-qubit Pauli group and the $2n$-qubit Bell basis. In this context, a generic basis element can be written as $\ket{P}\equiv \id\otimes P\ket{\id}$, where $\ket{\id}$ denotes the $2n$-qubit maximally entangled state, defined as $\ket{\id}=2^{-n/2}\sum_{i}^{d} \ket{i}\otimes\ket{i}$. Let us consider the state $\ket{\psi^*}\otimes\ket{\psi}$, it is not difficult to show that measuring in the Bell basis return outcome $P$ with probability $\tr(P\psi)^2/d$, with $\psi$ the density matrix associated to the state $\psi$. The probability to obtain $P$ is given by $|\braket{P|\psi^*\otimes\psi}|^2$
\ba
|\braket{P|\psi^*\otimes\psi}|^2&=&\braket{\id | \id\otimes P [\psi^* \otimes \psi] \id \otimes P|\id }\\ 
&=& \braket{\id|\id\otimes P\psi P\psi|\id}\\
&=&\tr(P\psi P\psi)/2^n=\tr(P\psi)^2/2^n
\ea
where we used the properties of the maximally entangled state that $O\otimes \id\ket{\id}=\id\otimes O^T\ket{\id}$, and that the state $\psi$ is pure. If instead of having access to $\ket{\psi^*}\otimes\ket{\psi}$ one has instead access to $\ket{\psi}\otimes\ket{\psi}$ the probability of sampling $P$ is given by $|\braket{\psi|P|\psi^*}|^2/2^n$. Through Bell sampling, one can learn a stabilizer state $\ket{\sigma}$ if provided access to $\ket{\sigma^*}\otimes\ket{\sigma}$. In this process, Bell basis measurements are performed $K$ times, and the resulting measurement outcomes yield a set of strings denoted as $T$. This set $T$ consists of $2n$-dimensional strings, each identifying a Pauli operator $P$ such that $P\in G_{\sigma}$, where $G_{\sigma}$ is the stabilizer group of $\ket{\sigma}$. Finding an $n$-dimensional basis for the $T$ is equivalent to finding a set of generators for the stabilizer group $G_{\sigma}$.

The potential failure of this process occurs when all the $K$ samples lie in a subspace of $T$ with dimension less than $n$. The probability that all the obtained samples fall within such a subspace is $2^{-K}$. By applying the union bound, we find that the probability of these samples lying in one of the $2^n$ $(n-1)$-dimensional subspaces of $T$ is $2^{-K+n}$. To mitigate this probability of failure, one can choose $K=2n$, resulting in an exponentially vanishing probability of failure. With the set of generators for $G_{\sigma}$ obtained, one can learn the phases by making a single-shot measurement of the expectation value of the learned Pauli generators.

In cases where access is limited to $\ket{\sigma}\otimes\ket{\sigma}$, rather than $\ket{\sigma^*}\otimes\ket{\sigma}$, it is still possible to learn the stabilizer state through a similar procedure. First, it is worth noting that for a stabilizer state $\ket{\sigma^*}=\bar{P}\ket{\sigma}$, where $P$ is a Pauli string consisting of $Z,\id$ operators. Consequently, the probability of sampling the Pauli operator $P$ can be rewritten as $|\braket{\sigma|P\bar{P}|\sigma}|^2/2^n$. Although $P$ does not stabilize $\ket{\sigma}$, the Pauli operator $P\bar{P}$ does it; otherwise, the sampling probability would be zero. Thus, when two Pauli operators, $P_1$ and $P_2$, are sampled, their product stabilizes $\ket{\sigma}$. This can be demonstrated straightforwardly by noting that both $P_1\bar{P}$ and $P_2\bar{P}$ stabilize $\ket{\sigma}$, and their product, $P_1\bar{P}P_2\bar{P}$, remains a Pauli operator that stabilizes $\ket{\sigma}$. Furthermore, it is proportional to $P_1P_2$, with the proportionality factor possibly being $\pm 1$ due to the commutation relations between $\bar{P}$ and $P_2$.

Thanks to this property, instead of sampling $K$ times, one only needs to sample $K+1$ times to determine the set of generators for $G_{\sigma}$, while maintaining the same failure probability.
\section{Efficient verification of containment}\label{app:cont}
In this section, we provide an alternative algorithm based on the notion of diagonalizer to efficiently verify the containment of a Pauli operator $P$ to the set $G_{\psi_t}\cup h_1G_{\psi_t}\cup\ldots\cup h_l G_{\psi_t}$. Let $\mathbb{Z}_n\subset\mathbb{P}_n$ the subgroup of the Pauli group generated by $\{\sigma_{i}^{z}\}_{i=1}^{n}$, with $\sigma_{i}^{z}=\operatorname{diag}(1,-1)$ on the $i$-th qubit. Then let $\mathcal{D}\in\mathcal{C}_n$ be a Clifford circuit such that $\tilde{G}_{\psi_t}\equiv\mathcal{D}G_{\psi_t}\mathcal{D}^{\dag}\subset \mathbb{Z}_n$. Notice that $\mathcal{D}$ can be found from a set of generators of $G_{\psi_t}$ in time $O(n^2)$~\cite{leone_learning_2022,oliviero_black_2022}. Define $\tilde{h}_i\equiv \mathcal{D}h_i\mathcal{D}^{\dag}$, which requires computational complexity $O(n^2l)$. To check whether $P\not\in G_{\psi_{t}}\cup h_1G_{\psi_t}\cup\ldots\cup h_{k}G_{\psi_t}$ is sufficient to check that $\mathcal{D}P\mathcal{D}^{\dag}\not\in \tilde{G}_{\psi_t}\cup \tilde{h}_i\tilde{G}_{\psi_t}\cup\ldots\cup \tilde{h}_l\tilde{G}_{\psi_t}$ that --  thanks to the local support of $\sigma_i^{z}s$ --  can be easily achieved by a qubit by qubit checking in time $O(nl)$. Consequently, the final runtime to verify if a Pauli operator $P$ belongs to $G_{\psi_t}\cup h_1G_{\psi_t}\cup\ldots\cup h_l G_{\psi_t}$ is $O(n^2k)$.

\section{Proofs}\label{app:proofs}
\subsection{Disjoint cosets.} \label{app:proofs_dc}
In this section, we will prove the disjointedness of the cosets. Let us consider $P\in S_{\psi_t}$ such that $P\not\in G_{\psi_t}$, then it is easy to notice that $\tr(Pg\psi_t)=\pm\tr(P\psi_t)$, and so $Pg\in S_{\psi_t} \, \forall g\in G_{\psi_t}$. To prove disjointedness let us consider two Pauli operators $P_{1},P_2\in S_{\psi_t}$ such that $P_{1},P_2\not\in G_{\psi_t}$, then if $P_{1}G_{\psi_t}\cap P_{2}G_{\psi_t}\neq 0$ would imply that $\exists g_{1},g_{2}\in G_{\psi_t}$ such that $P_{1}g_1=P_{2}g_{2}$. However, we would get that $P_{1}\in P_{2}G_{\psi_t}$, thanks to the group property of $G_{\psi_t}$, and thus $P_{1}G_{\psi_t}\equiv P_{2}G_{\psi_t}$ .

\subsection{Failure probability}\label{app:proof_fp}
Let us discuss the failure probability to sample a Pauli operator $P$ belonging to $G_{\psi_t}$. The event that has to be analyzed is the case where, sampled a Pauli $P\notin G$ after a set of measurements $M$ we are unable to distinguish it from a $P\in G_{\psi_t}$. Consequently, the failure probability is given by the joint probability that a measurement yields $1$ $M$ times and $P \notin G_{\psi_t}$. The probability that $P \notin G_{\psi_t}$ is given by $1-2^{-t}$, and denoting $p$ as the probability of obtaining $1$ ($p = \frac{1}{2}(\tr(h\psi_t)+1)$, where $h$ represents one bad generator), the failure probability is bounded by $(1-2^{-t})(h_{\max}/2+1/2)^M$.

\begin{comment}
{\em Finite resolution.---} One can proceed by induction. Given $\delta_1\ge (1-2^{-1/2})$, then let us pose $\delta_{t}\ge a_t $ where $a_t\equiv(1-2^{-1/2})^t$. For $t+1$ we have: $\delta_{t+1}\ge \min 2^{-1/2}||\tr[(\tilde{P}-\tilde{\tilde{P}})\psi_t]|- |\tr[(\tilde{Q}-\tilde{\tilde{Q}})\psi_t]||$. Where we used the fact that $\ket{\psi_{t+1}}=U_1\ket{\psi_t}$, where $U_1$ contains a single $T$-gate and thus $U_{1}^{\dag}PU_1=2^{-1/2}(\tilde{P}\pm \tilde{P})$. Noting that the smallest gaps between the expectation values scale of a factor $\sqrt{2}$, then $\delta_{t+1}\ge 2^{-1/2}(\sqrt{2}a_t-a_t)$ and one has $\delta_{t+1}\ge (1-2^{-1/2})a_t$ and thus $\delta_{t+1}\ge (1-2^{-1/2})^{t+1}$, which gives $b\simeq 1.77$.
\end{comment}

\subsection{\em Proof of Eq.~\eqref{probabilitypil}} 
Given the probability $\pi_{l}$ defined in Eq.~\eqref{probabilitypil2}, let us write explicitly the unity purity of the state $\psi_t$. From Eq.~\eqref{pgphy}, we obtain
\be
1=\frac{|G_{\psi_t}|}{d}\left(1+\sum_{i=1}^{l}\tr^2(h_{i}\psi_t)+\sum_{i=l+1}^{k}\tr^2(h_{i}\psi_t)\right)
\ee
Therefore, one can express $\pi_l$ as
\be
\pi_l=\frac{|G_{\psi_t}|}{d}\sum_{i=l+1}^{k}\tr^2(h_{i}\psi_t)\ge \frac{|G_{\psi_t}|}{d}(k-l)h^2_{\min}
\ee
which concludes the proof.
\subsection{Finite resolution}\label{proof_fr}
In this section, we present the proof of the finite resolution of the expectation values of Pauli operators for a $t$-doped stabilizer state. We will proceed as follows: we first bound $\min_{P\in S_{\psi_t}}|\tr(P\psi_t)|$ and then we bound the gap $\delta_t$, as the second will be just a trivial generalization of the first one. We will indeed find that
\be
\min_{P\in S_{\psi_t}}|\tr(P\psi_t)|\ge  \frac{1}{3\sqrt{2}^{t-1}}\left(1-\frac{1}{\sqrt{2}}\right)^{t}\label{minbound}
\ee
while
\be
\delta_t\ge  \frac{1}{6\sqrt{2}^{t-1}}\left(1-\frac{1}{\sqrt{2}}\right)^{t}\label{proofdeltabound}
\ee
Since a $t$-doped stabilizer state is built out from a $t$-doped Clifford circuit, we can alternatively bound $\min_P|\tr(C_{t}PC_{t}^{\dag}\sigma)|$ for $\sigma$ being an arbitrary stabilizer state. Let us look at the action of $C_t$ on a Pauli operator. First decompose $C_{t}=\prod_{i=1}^{t}C_{1}^{(i)}$ where $C_{1}^{(i)}$ is a $(t=1)$-doped Clifford circuit. Let us set up the following notation.
\be
C_{1}^{(1)}PC_{1}^{(1)\dag}=x_{(0)}P_{(0)}+\frac{x_{(1)}}{\sqrt{2}}P_{(1)}+\frac{x_{(2)}}{\sqrt{2}}P_{(2)}\label{1111}
\ee
where $x_{(0)},x_{(1)},x_{(2)}\in \{-1,0,+1\}$. Eq.~\eqref{1111} must be understood as: there is a choice of $x_{(0)},x_{(1)},x_{(2)}$ and the respective Pauli operators $P_{(0)},P_{(1)},P_{(2)}$ such that the l.h.s. is equal to the r.h.s. of Eq.~\eqref{1111}. Before generalizing to the generic $t$, it is useful to act again on $C_{1}^{(1)}PC_{1}^{(1)\dag}$ with $C_{1}^{(2)}$.
\ba
C_{1}^{(2)}C_{1}^{(1)}PC_{1}^{(1)\dag}C_{1}^{(2)\dag}&=&x_{(00)}P_{(00)}+
\frac{1}{\sqrt{2}}(x_{(10)}P_{(10)}+x_{(01)}P_{(01)}+x_{(20)}P_{(20)}+x_{(02)}P_{(02)})\nonumber\\&+&
\frac{1}{2}(x_{(11)}P_{(11)}+x_{(12)}P_{(12)}+x_{(21)}P_{(21)}+x_{(22)}P_{(22)})\label{2222}
\ea
where each variable $x_{(ij)}$ for $i=0,1,2$ can take values in $x_{(ij)}\in\{-1,0,+1\}$. As one can see, the subscript string $(ij)$ attached to each variable $x_{(ij)}$ reveals how many times a $T$-gate splits the Pauli operator $P$ in $2$ Pauli operator with the corresponding $\frac{1}{\sqrt{2}}$ factor. Again, Eq.~\eqref{2222} must be understood as there exists a choice of the variables $x_{(ij)}$ and the respective Pauli operators $P_{(ij)}$ for which the l.h.s. and the r.h.s. of Eq.~\eqref{2222} agrees. Now, that we set up the above general and powerful notation, we can easily generalize the action to $C_t$. We have the following
\be
C_{t}PC_{t}^{\dag}=\sum_{k=0}^{t}\sum_{\pi\in S_{k}}\frac{x_{\pi(\boldsymbol{y}_k)}}{\sqrt{2}^{k}}P_{\pi(\boldsymbol{y}_k)}\label{3333}\,.
\ee
In Eq.~\eqref{3333} above, we have defined a few elements. First of all, we defined the $t$-long string $\boldsymbol{y}_k$ with Hamming weight $k$ as
\be
\boldsymbol{y}_{k}=(\underbrace{1,1,\ldots,1}_{k},0,\ldots,0)
\ee
Next, we defined a set $S_{k}$ of operations $\pi$ that act on $\boldsymbol{y}_{k}$. $S{k}$ is the set containing all the permutations of the $k$ $1$s in $\boldsymbol{y}_k$ into $t$ spots, combined with the operation that transforms $1\leftrightarrow2$, in accordance with Eq.~\eqref{2222}. Let's illustrate this with an example. Set $t=2$ and $k=1$, so $\boldsymbol{y}_{1}=(10)$. All the possible permutations of $(10)$, combined with the operation $1\leftrightarrow 2$, result in the strings $(10),(01),(20),(02)$. Similarly, for $t=2$ and $k=2$, $\boldsymbol{y}_2=(11)$ and the set of operations in $S_{2}$ returns $(11),(22),(21),(12)$.

It is useful to count the number of operations within $S_{k}$ for fixed $k$. The set $S_{k}$ is the combination of $\binom{t}{k}$ many ways to permute $\boldsymbol{y}_k=(1,1,\ldots, 1,0,\ldots, 0)$ times the $2^k$ different choices of either $1$ or $2$ at any site. The above simple counting thus returns:
\be
\sum_{\pi\in S_{k}}=2^{k}\binom{t}{k}
\ee
From Eq.~\eqref{3333}, we can formally compute the expectation value of $C_{t}PC_{t}^{\dag}$ with a generic stabilizer state $\sigma$ and get
\be
\tr(C_{t}PC_{t}^{\dag}\sigma)=\sum_{k=0}^{t}\sum_{\pi\in S_{k}}\frac{\tilde{x}_{\pi(\boldsymbol{y}_k)}}{\sqrt{2}^{k}}
\ee
where we defined the variables $\tilde{x}_{\pi(\boldsymbol{y}_k)}:=x_{\pi(\boldsymbol{y}_k)}\tr(P_{\pi(\boldsymbol{y}_k)}\sigma)\in\{-1,0,+1\}$ because $\tr(P_{\pi(\boldsymbol{y}_k)}\sigma)\in\{-1,0,+1\}$. Now, we set up all the necessary notation to prove Eq.~\eqref{minbound}. 

We are interested in computing the minimum achievable value for $\tr(C_{t}PC_{t}^{\dag}\sigma)$. Let us first multiply both sides for $\sqrt{2}^{t}$. We thus get
\ba
\sqrt{2}^{t}\tr(C_{t}PC_{t}^{\dag}\sigma)=\sum_{k=0}^{t}\sum_{\pi\in S_{k}}\sqrt{2}^{t-k}\tilde{x}_{\pi(\boldsymbol{y}_k)}=
\sum_{l=0}^{t}\sum_{\pi\in S_{t-l}}\sqrt{2}^{l}\tilde{x}_{\pi(\boldsymbol{y}_{t-l})}\nonumber
\ea
where in the second equality, we defined $l=t-k$. Let us set $t$ to be even and split odd and even terms in the sum
\ba
\sum_{l=0}^{t}\sum_{\pi\in S_{t-l}}\sqrt{2}^{l}\tilde{x}_{\pi(\boldsymbol{y}_{t-l})}=\sum_{l=0}^{t/2}2^l\sum_{\pi\in S_{t-2l}} \tilde{x}_{\pi(\boldsymbol{y}_{t-2l})}+
\sqrt{2}\sum_{l=0}^{t/2-1}2^l\sum_{\pi\in S_{t-(2l+1)}} \tilde{x}_{\pi(\boldsymbol{y}_{t-(2l+1)})}\nonumber
\ea
Define the following function of $t$
\ba
A(t)&:=&\sum_{l=0}^{t/2}2^l\sum_{\pi\in S_{t-2l}} \tilde{x}_{\pi(\boldsymbol{y}_{t-2l})}\\
B(t)&:=&\sum_{l=0}^{t/2-1}2^l\sum_{\pi\in S_{t-(2l+1)}} \tilde{x}_{\pi(\boldsymbol{y}_{t-(2l+1)})}
\ea
Note that $A(t),B(t)\in\mathbb{Z}$, i.e. they are positive and negative natural numbers, for any $t=2t^{\prime}$ for $t\in\mathbb{N}$. We can thus write 
\be
\sqrt{2}^{t}|\tr(C_{t}PC_{t}^{\dag}\sigma)|=|A(t)+\sqrt{2}B(t)|=|B(t)|\left|\sqrt{2}+\frac{A(t)}{B(t)}\right|\label{4444}
\ee
Therefore, the lower bound deals with the approximation of the algebraic number $\sqrt{2}$ by a rational number $A(t)/B(t)$. To make it explicit we can lower bound the r.h.s. of Eq.~\eqref{4444} as
\be
\sqrt{2}^{t}|\tr(C_{t}PC_{t}^{\dag}\sigma)|\ge |B(t)|\left|\sqrt{2}-\frac{|A(t)|}{|B(t)|}\right|\label{6666}
\ee
and we can invoke the Liouville Theorem (see Ref.~\cite{murty_liouville_2014}) of approximating a algebraic number $\alpha$ with two rational numbers $p,q\in\mathbb{Q}$ that reads: there exist a constant $c(\alpha)$ independent from $p,q$ such that 
\be
\left|\sqrt{2}-\frac{p}{q}\right|\ge \frac{c(\alpha)}{q^{D}}\label{5555}
\ee
where $D$ is the degree of the algebraic number $\alpha$. In the case of $\alpha=\sqrt{2}$ we have $D=2$ because $\sqrt{2}$ corresponds to the solution to the irreducible polynomial $z^2-2=0$ which has degree $2$, and $c(\sqrt{2})=\frac{1}{6}$~\cite{murty_liouville_2014}. Applying Eq.~\eqref{5555} to Eq.~\eqref{6666}, we thus get
\be
|\tr(C_{t}PC_{t}^{\dag}\sigma)|\ge \frac{1}{6|B(t)|\sqrt{2}^{t}}
\ee
We are left to find an upper bound to $B(t)$. We proceed with the following equality
\ba
|B(t)|&=&\left|\sum_{l=0}^{t/2-1}2^l\sum_{\pi\in S_{t-(2l+1)}} \tilde{x}_{\pi(\boldsymbol{y}_{t-(2l+1)})}\right|\le \sum_{l=0}^{t/2-1}2^l\sum_{\pi\in S_{t-(2l+1)}} |\tilde{x}_{\pi(\boldsymbol{y}_{t-(2l+1)})}|\nonumber\\&=&\sum_{l=0}^{t/2-1}2^l\sum_{\pi\in S_{t-(2l+1)}} = \sum_{l=0}^{t/2-1}2^l2^{t-(2l+1)}\binom{t}{2l+1}\nonumber\\&=&\frac{1}{\sqrt{8}}2^t\left[\left(1+\frac{1}{\sqrt{2}}\right)^t-\left(1-\frac{1}{\sqrt{2}}\right)^t\right]\le \frac{1}{\sqrt{8}}\left(1-\frac{1}{\sqrt{2}}\right)^{-t}\label{Bbound}
\ea
where in the second equality, we used the fact that $\tilde{x}_{\pi(\boldsymbol{y}_{t-(2l+1)})}\in\{-1,0,+1\}$ and in the last inequality, we used the fact that 
\be
\frac{1}{\sqrt{8}}2^t\left[\left(1+\frac{1}{\sqrt{2}}\right)^t-\left(1-\frac{1}{\sqrt{2}}\right)^t\right]\le\frac{1}{\sqrt{8}}2^t\left(1+\frac{1}{\sqrt{2}}\right)^t= \frac{1}{\sqrt{8}}\left(1-\frac{1}{\sqrt{2}}\right)^{-t}
\ee 
An analogous procedure with $t=2t^{\prime}+1$ with $t^{\prime}$ leads to the same exact bound. Therefore for any $t\in\mathbb{N}$, we find
\be
|\tr(C_{t}PC_{t}^{\dag}\sigma)|\ge \frac{\sqrt{8}}{6}\left(\frac{1}{\sqrt{2}}-\frac{1}{2}\right)^{t}
\ee
Now, let us turn to analyze the gap $\delta_t\equiv\tr[(P-P^{\prime})\psi_t]$. Using the same notation as before, we can write the adjoint action of $C_{t}$ on $P-P^{\prime}$ as follows
\be
C_{t}(P-P^{\prime})C^{\dag}_t=\sum_{k=0}^{t}\sum_{\pi\in S_{k}}\left(\frac{x_{\pi(\boldsymbol{y}_k)}}{\sqrt{2}^{k}}P_{\pi(\boldsymbol{y}_k)}+\frac{x^{\prime}_{\pi(\boldsymbol{y}_k)}}{\sqrt{2}^{k}}P^{\prime}_{\pi(\boldsymbol{y}_k)}\right)
\ee
and therefore, by repeating the same procedure, we can define 
\ba
A^{\prime}(t)&:=&\sum_{l=0}^{t/2}2^l\sum_{\pi\in S_{t-2l}} \tilde{x}_{\pi(\boldsymbol{y}_{t-2l})}+\tilde{x}^{\prime}_{\pi(\boldsymbol{y}_{t-2l})}\\
B^{\prime}(t)&:=&\sum_{l=0}^{t/2-1}2^l\sum_{\pi\in S_{t-(2l+1)}} \tilde{x}_{\pi(\boldsymbol{y}_{t-(2l+1)})}+\tilde{x}^{\prime}_{\pi(\boldsymbol{y}_{t-(2l+1)})}
\ea
and write the gap as
\be
\delta_t=\frac{1}{\sqrt{2}^t}|A^{\prime}(t)+\sqrt{2}B^{\prime}(t)|\ge \frac{|B^{\prime}(t)|}{\sqrt{2}^t}\left|\sqrt{2}-\frac{|A^{\prime}(t)|}{|B^{\prime}(t)|}\right|\ge \frac{1}{6\sqrt{2}^t|B^{\prime}(t)|}
\ee
Following the inequalities in Eq.~\eqref{Bbound}, one can find
\be
|B^{\prime}(t)|\le \frac{1}{\sqrt{2}}\left(1-\frac{1}{\sqrt{2}}\right)^{-t}
\ee
which recovers the desired result in Eq.~\eqref{proofdeltabound}
\be
\delta_t\ge \frac{\sqrt{2}}{6}\left(\frac{1}{\sqrt{2}}-\frac{1}{2}\right)^{t}
\ee
\section{Sampling the stabilizer group from Bell measurements}\label{App:sampling}
\begin{lemma}\label{lemmaexistence}
Given a quantum state $\ket{\omega}$ on $q$-qubits, there exist a Pauli operator $\bar{P}$ such that $|\braket{\omega|\bar{P}|\omega^{*}}|\ge 2^{-q}$. 
\begin{proof}
To show this, let us assume the contrapositive, i.e. $\forall P\in\mathbb{P}_q$ then $|\braket{\omega|P|\omega^{*}}|< 2^{-q}$. First of all, notice that it always holds 
\be
\sum_{P\in\mathbb{P}_q}|\braket{\omega|P|\omega^{*}}|^2=\sum_{P\in\mathbb{P}_q}\tr(\omega P\omega^{*}P)=2^{q}
\ee
However, sticking to the hypothesis, if we sum up the squares we get
\be
\sum_{P\in\mathbb{P}_q}|\braket{\omega|P|\omega^{*}}|^2<2^{-q}\sum_{P\in\mathbb{P}_q}<2^{q}
\ee
and this is a contradiction.
\end{proof}
\end{lemma}
In this section, we shall show that given $P,P^{\prime}\sim\widetilde{\Xi}_{\psi_t}$ drawn from $\widetilde{\Xi}_{\psi_t}$, then their product $PP^{\prime}$ belongs to the stabilizer group $G_{\psi_t}$ with probability greater than $2^{-6t}$. Formally
\be
\underset{P,P^{\prime}\sim\widetilde{\Xi}_{\psi_t}}{\operatorname{Pr}}(PP^{\prime}\in G_{\psi_t})\ge \frac{1}{2^{6t}}\label{statementproof}
\ee
First of all, let us refresh some basis about stabilizer states. Consider a stabilizer state $\ket{\sigma}$ and its conjugate state in the computational basis $\ket{\sigma^{*}}$. In Ref.~\cite{montanaro_learning_2017}, it has been shown that there exist a Pauli operator $P_{\sigma}$ such that 
\be
P_{\sigma}\ket{\sigma}=\ket{\sigma^{*}}
\label{pzstab}
\ee
which can be understood at the level of the stabilizer group in the following way. Let $G_{\sigma}=\braket{g_{1},\ldots, g_{n}}$ the stabilizer group of $\ket{\sigma}$. The stabilizer group of $\ket{\sigma^{*}}$ is, of course, given by $G_{\sigma^{*}}=\braket{g_{1}^{*},\ldots, g_{n}^{*}}$. Therefore, thanks to Eq.~\eqref{pzstab}, there exist $P_{\sigma}$ such that
\be
P_{\sigma}g_{i}P_{\sigma}=g_{i}^{*},\quad \forall g_{i}\in G_{\sigma}
\ee
Consider now the projector $\Pi_{\psi_t}:=\prod_{i=1}^{m}\frac{\id+\phi_{g_j}g_j}{2}$ onto the stabilizer group $G_{\psi_t}$. Since $G_{\psi_t}$ can be completed to a maximal stabilizer group $\widetilde{G_{\psi_t}}$, i.e. $|\widetilde{G_{\psi_t}}|=d$, there exists $P_{\psi_t}$ such that
\be
P_{\psi_t}\Pi_{\psi_t}P_{\psi_t}=\Pi_{\psi_t}^{*}
\ee
where $\Pi_{\psi_t}^{*}=\prod_{i=1}^{m}\left(\frac{\id+\phi_{g_j}g_j^{*}}{2}\right)$. On the state vector level, the action of $P_{\psi_t}$ on $\ket{\psi_t^{*}}$ results in $\ket{\psi_{t}^{*\prime}}=P_{\psi_t}\ket{\psi_t^{*}}$ with a density matrix representation equal to
\be
\st{\psi_{t}^{*\prime}}=\left(\frac{1}{2^t}\sum_{i=0}^{k}\tr(h_i\psi_t^{*})h_i\right)\Pi_{\psi_t}\label{psitstarprime}
\ee
i.e. having the same projector $\Pi_{\psi_t}$ of $\ket{\psi_t}$. To demonstrate the validity of Eq.~\eqref{psitstarprime}, is sufficient to compute the overlap $\tr(\Pi_{\psi_t} P_{\psi_t}\psi_{t}^{*}P_{\psi_t})=1$. Thanks to Eq.~\eqref{psitstarprime}, given $\mathcal{D}_{\psi_t}$ being the diagonalizer associated to $\ket{\psi_t}$, one has that 
\be
\mathcal{D}_{\psi_t}\ket{\psi_{t}^{*\prime}}=\ket{0}^{\otimes m}\otimes \ket{\phi^{*}}
\ee
The last step before proving the statement comes from the result of Lemma~\ref{lemmaexistence} that ensures the existence of a Pauli operator $\bar{P}$ such that $|\braket{\phi|\bar{P}|\phi^{*}}|\ge 2^{-(n-m)}$. Therefore, let us prove the statement, i.e. that the probability that Bell measurements sample a Pauli $P$ such that $PP_{\psi_t}\mathcal{D}^{\dag}\bar{P}\mathcal{D}\in G_{\psi_t}$ is greater than $d^{-1}8^{-t}$. Starting Eq.~\eqref{probabilityreal}, we have 
\ba
|\braket{\psi_{t}|P|\psi_{t}^{*}}|&=& |\braket{\psi_{t}|PP_{\psi_t}|\psi_{t}^{*\prime}}|\nonumber\\&=&|\braket{0^m\otimes\phi|\mathcal{D}PP_{\psi_t}\mathcal{D}^{\dag}|0^m\otimes \phi^{*}  }|\nonumber\\&=&
|\braket{0^m\otimes\phi|\mathcal{D}PP_{\psi_t}\mathcal{D}^{\dag}\bar{P}\bar{P}|0^m\otimes \phi^{*}  }|\nonumber
\ea
Let us impose $PP_{\psi_t}\mathcal{D}^{\dag}\bar{P}\mathcal{D}\in G_{\psi_t}$; therefore, we have the following bound:
\ba
|\braket{\psi_{t}|P|\psi_{t}^{*}}|&=&|\braket{0^m\otimes\phi|\bar{P}|0^m\otimes \phi^{*}  }|\nonumber\\&=&|\braket{\phi|\bar{P}|\phi^{*}}|\ge 2^{-(n-m)}
\ea
And therefore, we can formally write
\be
\operatorname{Pr}(PP_{\psi_t}\mathcal{D}^{\dag}\bar{P}\mathcal{D}\in G_{\psi_t})\ge \frac{|G_{\psi_t}|}{d}\frac{1}{2^{2(n-m)}}\ge \frac{1}{8^t}\label{111111}
\ee
For fixed $P_{\psi_t}$ and $\bar{P}$, one has that the probability that two samples Pauli operators $P,P^{\prime}$ obeys the property in Eq.~\eqref{111111} is $\frac{1}{{2}^{6t}}$, and one has that the probability that the product $PP^{\prime}\in G_{\psi_t}$ is, by a simple shift for $P_{\psi_t}\mathcal{D}^{\dag}\bar{P}\mathcal{D}$, is the one in Eq.~\eqref{statementproof}.
\begin{comment}
\section{Average case complexity}
In this section, we discuss the average case complexity of our algorithm. The set of states we are interested in is the set of states with bounded stabilizer nullity $\nu(\ket{\psi})$ that we define rigorously as
\be
S_{\nu}:=\{\ket{\psi}\in\mathcal{H}\,|\, \nu(\ket{\psi})\le t\}
\ee
First notice that $t$-doped stabilizer states are included in $S_{\nu}$. Then, we have that for every state $\ket{\psi}\in S_{\nu}$, then the state $\ket{\psi}$ admit the following decomposition~\cite{leone_learning_2022}
\be
\ket{\psi}=\mathcal{D}(\ket{0}^{\otimes n-t}\otimes \ket{\phi_t})\label{decompositionstate}
\ee
where $\mathcal{D}$ is a Clifford operator, known as \textit{diagonalizer}, and $\ket{\phi_t}$ is a $t$-qubit quantum states. Thanks to the property~\eqref{decompositionstate}, the most natural distribution that brings together all the states in $S_{\nu}$ is $\pi_{S_{\nu}}=\haar(\mathcal{C}_n)\circ \haar(\mathcal{U}_t)$, i.e. uniform average over the Block sphere of $t$-qubit and uniform average of $\mathcal{D}$. That said, we have the following result
\be
\operatorname{Pr}(|h_{\min}-2^{-t/2}|\ge2^{-bt})\ge1-22^{-2^{1-2bt}+2\log t}
\ee
\end{comment}

\end{document}